\documentclass[twocolumn,showpacs,preprintnumbers,amsmath,amssymb]{revtex4}
\usepackage{amssymb}

\usepackage{graphicx}
\usepackage{dcolumn}
\usepackage{bm}
\usepackage{color}

\begin{document}

\preprint{}


\title{Controllable pulse patterns in fiber lasers}

\author{Xingliang Li}
\author{Shumin Zhang}
\email{zhangsm_sd@126.com}
\author{Jingmin Liu}
\author{Zhenjun Yang}

\address{College of Physics, Hebei Advanced Thin Films Key Laboratory, Hebei Normal University,
Shijiazhuang 050024, China}



\begin{abstract}
An all-optical pulse power editing (PPE) technique is reported. Using the PPE technique, pulses with different peak powers are output and directed to the positive or reverse saturable absorption (SA) range of the saturable absorber. Further, under the combined action of the PPE technique and SA, four pulse patterns including dissipative soliton (DS), DS molecules, soliton compounds composed of DS and noise-like pulse (NLP), and pure NLP are controllably generated in fiber lasers. The results are conducive for developing advanced DS lasers and can further clarify the onset of pulse dynamic patterns.
\end{abstract}

\pacs{42.65.Re, 42.55.Wd} 

\maketitle


\section{Introduction}
The observation, prediction, and production of nonlinear dynamic patterns in dissipative systems are ubiquitous in many scientific fields including hydrodynamic systems \cite{MCCrossP-1}, plasma physics \cite{YTsaiY-2}, chemistry \cite{GRucknerR-3}, biology \cite{CBlanch-MercaderJ-4}, and fiber optics \cite{LGWrightD-5}. As the fiber is highly nonlinear, the fiber lasers can provide an excellent framework for the generation of pulse dynamic patterns such as dissipative solitons (DSs) \cite{PGreluN-6, SHongF-7}, DS molecules \cite{KKrupa-8,XLiuX-9,GHerinkF-10,BOrtacA-11,AZavyalovR-12,AKomarovF-13}, noise-like pulses (NLPs) \cite{MHorowitzY-14,SKerenM-15,LMZhao-16,SKobtsevS-17,AFJRunge-18,WchangJ-19,MSuzukiR-20,XWangA-21,DVChurkin-22,XliS-23}, and rogue waves \cite{CLecaplainPh-24,CLecaplainPh-25}.

Pulse propagation in fibers can be well modeled by the nonlinear Schr\"{o}dinger equation (NLSE) \cite{GPAgrawal-26}. Considering the amplitude modulation effects in a fiber laser cavity, certain terms are added to the NLSE, and the resulting differential equation is called a cubic-Ginzburg-Landau equation (CGLE). The solution of the CGLE can be categorized as a DS with large net normal fiber cavity dispersion \cite{HAHaus-27}. Specifically, the energy, profile, and chirp of the DS pattern are predetermined by the equation parameters, rather than the initial conditions \cite{YSongX-28}. Hence, an extended exploration in the parameter space can reveal numerous DS pattern varieties. Generally, the pattern variety is considerably wider than can be imagined and approximated \cite{PGreluN-6}. However, there is no effective way to manipulate the cavity parameter space for the controlled and orderly production of different pulse  patterns, as yet.

Fiber nonlinearities involve the generation of new spectral components and pulse breaking \cite{GPAgrawal-26}. The diversity of the pulse pattern is generally considered to be due to the accumulated remarkably large nonlinear phase \cite{PGreluN-6,FWWiseA-29}. On the other hand, fiber dispersion plays a critical role in the temporal broadening or compression of optical pulses. For instance, using the considerable dispersion, dispersive Fourier transformation can map the single-shot spectrum of a pulse pattern to a low power broadening temporal waveform, enabling the real-time exploration of the ultrafast transient process \cite{PRyczkowskiM-30, KGodaB-31}. Amplifying the time-stretch pulse caused by medium amounts of dispersion using an ingenious approach enables higher power pulse generation \cite{DStricklandG-32}. Therefore, the pulse power can be effectively edited by combining the laser gain and fiber length that is dependent on the pulse amplification and the dispersion broadening. Furthermore, a bandwidth-limited spectral filter can be used as the standardized spectral component selector enabling further pulse power editing, quantitatively and versatilely. The pulse power editing, in turn, further affects its nonlinear accumulation in the cavity.

In addition to the pulse patterns caused by the excessive nonlinear phase shift, DSs and NLPs can also be generated by managing the saturable absorption (SA) of the nonlinear polarization rotation (NPR), as reported in \cite{YJeongL-33, ZChengH-34}. Cheng et al. also numerically demonstrated that square-shaped emission is caused by reverse saturable absorption (RSA) with a modulation depth of 1.5\% in a Yb-doped fiber laser \cite{ZChengH-35}. Subsequently, Li et al. experimentally proved the RSA of NPR \cite{XliS-23}, and explicitly defined the critical saturation power (CSP), and the positive and reverse SA ranges of NPR. However, the RSA of NPR can also be circumvented by employing dispersion-induced pulse broadening within the cavity \cite{BGBaleS-36}, and reducing the peak power while maintaining a single peak DS envelope. The question arises as to whether the pulse patterns can be controlled to produce in the laser cavity by managing the peak power of the pulse entering or bypassing the RSA region of saturable absorber. This issue formed the initial motivation for this Letter.

In this work, combining spectral filtering, pulse amplification, fiber dispersion and nonlinearity, a simple all-optical PPE link is presented. By combining the PPE technique and the SA effect of the NPR, pulse patterns including DSs, DS molecules, soliton compounds composed of DS and NLP, and pure NLP are controllably and orderly produced in fiber-optic media cavitys for the first time. This can assist in understanding the mechanism of pulse patterns generation, and define an effective way to manipulate the DS cavity parameters in an orderly manner.

\section{Concept description}
Qualitatively, the round-trip cavity model comprises an all-optical PPE link and a lumped NPR as shown in \textbf{Figure 1}. In the inset of Figure 1, the NPR power transmission curve which is the product of the instantaneous pulse power and the NPR transmissivity versus the instantaneous power is depicted \cite{XliS-23}. The CSP of the first mode-locked region is 692.0 W. Nonlinear transmission occurs due to an intracavity polarizer that transforms the nonlinear evolution of the polarization state into intensity-dependent losses.

When the sum of the lengths of single-mode fibers ($\rm{SMF_{1}}$ and $\rm{SMF_{2}}$) is fixed in Figure 1, the same output pulse from the band-pass filter (BPF) will experience the same dispersion effect, extending the pulse duration according to the pulse broadening formula, \emph{$\Delta$}\emph{T} = \emph{DL}$\Delta\lambda$ ($\emph{D}$ is the dispersion parameter, $\emph{L}$ is the fiber length, and \emph{$\Delta\lambda$} is the 3-dB spectral width). As the original pulse peak power is lower, dispersion alone affects the pulse duration when the pulse propagates along $\rm{SMF_{1}}$; however, when the pulse passes through the gain fiber, its power is amplified. When the pulse with higher peak power propagates through $\rm{SMF_{2}}$, nonlinearity also affects the pulse duration. In a normal-dispersion fiber, the combined effect of dispersion and nonlinearity causes increased pulse broadening, compared to the individual effect of dispersion. In addition, the pulse peak power reduces because of the combined effect of dispersion and nonlinearity in $\rm{SMF_{2}}$. Hence, the longer the length of $\rm{SMF_{2}}$, the lower is the pulse peak power. As an example, the upper diagram in \textbf{Figure 2a} shows that the PPE link can be used to edit a pulse with a peak power of 246.8 W, at position B. After setting the other parameters constant, the longer $\rm{SMF_{2}}$ is swapped with $\rm{SMF_{1}}$, and a pulse with a peak power of 799.9 W is edited as depicted in the lower-right diagram of Figure 2b.
\begin{figure}[htbp]
\centering\includegraphics[width=6.2cm]{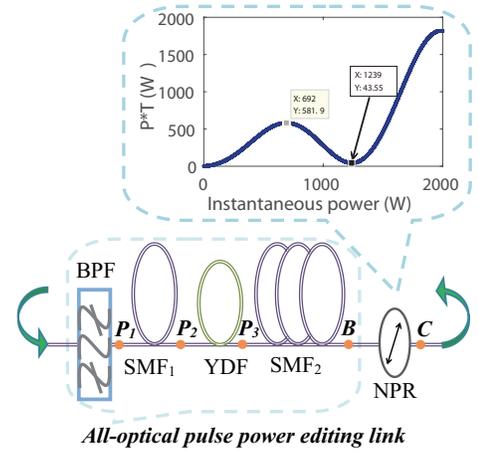} \caption{Conceptual schematic of a fiber laser comprising an all-optical pulse power editing (PPE) link and a lumped NPR. The NPR transmittance curve (inset) is the product of the instantaneous pulse power and NPR transmissivity versus the instantaneous power [Inset parameters used in the simulation: $\Delta\varphi_0$ = 0.05 $\pi$, $\delta\lambda/\lambda_s$ = 0, \emph{L}/\emph{L}$_b$ = 80, $\theta$ = 0.263 $\pi$, $\psi$ = 0.177 $\pi$].}\label{FL-Fig1}
\end{figure}

\begin{figure}[htbp]
\centering\includegraphics[width=8.5 cm]{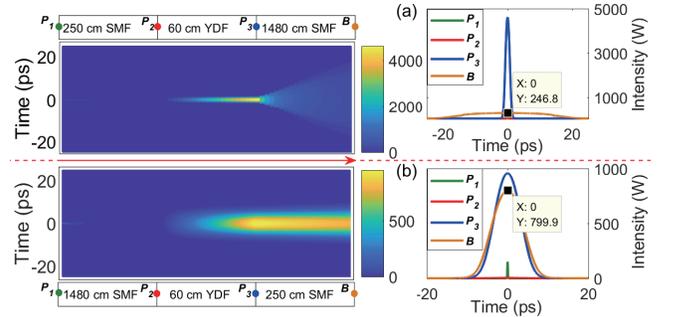} \caption{Two sample pulse evolution processes with the shorter $\rm{SMF_{1}}$ and the longer $\rm{SMF_{2}}$ (upper diagram), and the transposition of $\rm{SMF_{1}}$ and $\rm{SMF_{2}}$ in the pulse power edit link (lower diagram).}\label{FL-Fig2}
\end{figure}

It can be observed that when different lengths are selected for $\rm{SMF_{1}}$ and $\rm{SMF_{2}}$ with a fixed sum of fiber lengths, different pulse output peak power can be obtained; hence, we call this an all-optical PPE technique. Furthermore, when the peak power of the output pulse at position B on the PPE link is in the range of 0 - 692.0 W, as shown in the inset of Figure 1, the NPR provides a positive SA effect. On the contrary, when this value is located in the 692.0 - 1239.0 W range, the NPR provides a RSA effect. When this value slightly exceeds the CSP of 692.0 W, the NPR offers an absolutely positive SA effect and a small portion of the RSA effect. Based on the above, combining the PPE technique and the SA, different pulse dynamic patterns can be produced controllably in all-normal-dispersion fiber lasers.
\section{Numerical simulation}
Pulse propagation along the fiber sections in the presence of higher-order-dispersion and nonlinear effects is modeled using the generalized nonlinear Schr\"{o}dinger equation \cite{XLiS-37,JMDudleyS-38,ZZHUT-39}. The models of the amplitude modulation terms including NPR, gain can be found in Refs. \cite{XliS-23} and \cite{XLiS-37}. The PPE link comprises an 8-nm BPF with a central wavelength of 1030 nm, a segment of $\rm{SMF_{1}}$, a 60-cm-long YDF, and another segment of $\rm{SMF_{2}}$, in order. The group-velocity dispersion of the SMFs and YDF is assumed to be the same at 23 $\rm{ps^2}$/km, the nonlinear coefficients of the SMFs and YDF are 4.68 $\rm{W^{-1}}$$\rm{km^{-1}}$ and 9.3 $\rm{W^{-1}}\rm{km^{-1}}$, respectively, the gain bandwidth of the YDF is 45 nm, and the gain saturation energy is 1441 pJ. The linear cavity loss is 22\%. The simulation results demonstrate that different pulse dynamic patterns can be controllably generated in the YDF laser with a fixed laser cavity length by distributing the lengths of $\rm{SMF_{1}}$ and $\rm{SMF_{2}}$ alone, as shown in \textbf{Table 1} and \textbf{Figures 3-7}.

\begin{table}[]
\caption{\label{tab:Table}%
Controllably generated pulse patterns in fiber lasers. (DS: dissipative soliton, NLP: noise-like pulse)}
\begin{tabular}{|l|l|l|l|l|l|}
\hline
\begin{tabular}[c]{@{}l@{}}$\rotatebox{270}{Simulation}$\end{tabular}& \begin{tabular}[c]{@{}l@{}}$\rm{SMF_{1}}$\\(cm)\end{tabular} & \begin{tabular}[c]{@{}l@{}}$\rm{SMF_{2}}$\\(cm)\end{tabular} & \begin{tabular}[c]{@{}c@{}}$\rm{SMF_{1}}$\\+\\$\rm{SMF_{2}}$ \\(cm)\end{tabular} & \begin{tabular}[c]{@{}c@{}}$\rm{g_{0}}$ (dB)\\(pump\\ power)\end{tabular} & \begin{tabular}[c]{@{}l@{}}Pulse pattern\\(Figures)\end{tabular} \\ \hline
1   & 480                                             & 1250                                             & 1730                                             & 95.0                & \begin{tabular}[c]{@{}l@{}}Singe DS\\\textbf{Figure 3}\end{tabular}  \\ \hline
2   & 680                                             & 1050                                             & 1730                                             & 95.0                 & \begin{tabular}[c]{@{}l@{}}DS molecules\\\textbf{Figure 4}\end{tabular}  \\ \hline
3   & 880                                             & 850                                              & 1730                                             & 95.0                  & \begin{tabular}[c]{@{}l@{}}Soliton\\ compounds\\\textbf{Figure 5}\end{tabular}   \\ \hline
4   & 1480                                             & 250                                             & 1730                                            & 95.0                  & \begin{tabular}[c]{@{}l@{}}NLP\\\textbf{Figure 6}\end{tabular}   \\ \hline
\end{tabular}
\end{table}

\subsection{Single DS}
When the pulse peak power is lesser than the CSP before the pulse passes through the NPR, an absolutely positive SA is presented by the NPR, and the laser generates typically steady-state DSs \cite{XliS-23}. \textbf{Figure 3} shows the evolution of critical-state DSs in the YDF laser, when g$_0$ = 95 dB (Simulation 1 in Table 1). In this case, the peak power of the output pulse from the PPE link is 854.5 W, which exceeds the CSP value before the pulse passes through the NPR over consecutive round trips. In this case, the NPR presents absolutely positive SA and a small portion of RSA. Therefore, the pulse is in the form of a double peak after NPR at position C, as shown in Figure 3. The pulse is restored to a single peak after further shaping by the BPF. The original pulse continually circulates in the cavity and undergoes shaping in the PPE link, NPR, and filter, and finally, a stable DS is produced.

\subsection{DS molecules}
The lengths of $\rm{SMF_{1}}$ and $\rm{SMF_{2}}$ are varied from 480 to 680 cm and 1250 to 1050 cm, respectively, in the PPE link with a fixed g$_0$ and SMF length sum. The pulse evolution up to stabilization involves less than 55 round trips; the peak power \emph{P}$_p$ of the pulse gradually increases at position B and the pulse width T remains substantially constant because of the lower dispersion effect, as indicated by the arrow direction in the rightmost time evolution of \textbf{Figure 4} (cavity position after the filter). This pulse evolution suggests that the self-phase modulation (SPM) effect dominates the GVD effect because $\gamma$\emph{P}$_p$T$^2$$\gg$$\beta_2$ ($\gamma$ is the fiber nonlinear parameter, \emph{P}$_p$ is the peak power of the incident pulse, T is pulse width, and $\beta_2$ is the GVD parameter) at least during the initial stages of pulse evolution. The physical origin of the small pulse envelope near the arrows on the leftmost pulse evolution of Figure 4 (cavity position B) is related to optical wave breaking \cite{WJTomlinsonR-40} and YDF dispersion. The formed small pulse envelopes are subjected to positive SA while propagating along the cavity because of their lower peak power, until the laser system reaches steady state, and bound DSs (DS molecules) are finally formed. It is to be noted that because the laser is operated in the normal GVD ($\beta_2$ $>$ 0), the red-shifted light near the leading edge travels faster, whereas the blue-shifted light located near the trailing edge travels slower; when the red-shifted light overtakes the blue-shifted one in the forward tail of the pulse, interference occurs and interference fringes are formed, as depicted in the spectra of Figure 4.

It is worth mentioning that the edited pulse peak power at position B here is 948.1 W at the 44th round trip, which is higher than the CSP of 692.0 W; i.e. more instantaneous power enters into the NPR RSA range. The obvious reduction of the pulse peak power after passing through position C verifies the effectiveness of pulse power editing.

\begin{figure}[htbp]
\centering\includegraphics[width=8.2 cm]{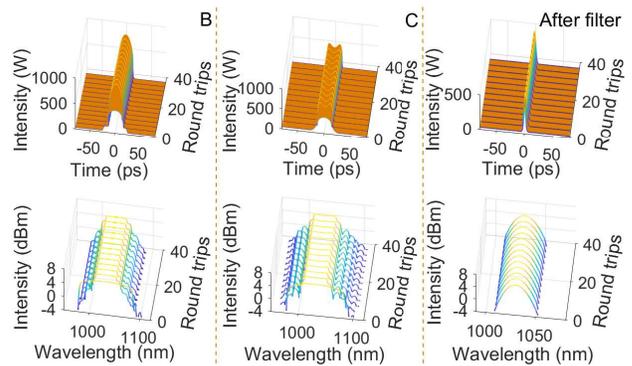} \caption{Calculated single steady-state DS evolution in the time (up) and spectral (down) domains for cavity positions B, C, and after the filter.}\label{FL-Fig3}
\end{figure}

\begin{figure}[htbp]
\centering\includegraphics[width=8.2cm]{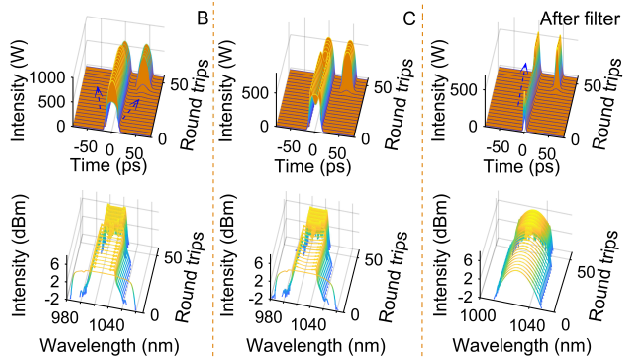} \caption{Calculated steady-state DS molecules evolution in the time (up) and spectral (down) domains for cavity positions B, C, and after the filter.}\label{FL-Fig4}
\end{figure}

\subsection{Soliton compounds}
The length of $\rm{SMF_{1}}$ is further increased from 680 cm to 880 cm, whereas that of $\rm{SMF_{2}}$ is decreased from 1050 cm to 850 cm, in the PPE link with a fixed g$_0$ and SMF length sum (Simulation 3 in Table 1). From the above analysis, the shorter the $\rm{SMF_{2}}$ length, the higher is the peak power of the output pulse from the PPE link. Optical wave breaking induced by the SPM effect in the YDF \cite{GPAgrawal-26} and fine-structure oscillations in the pulse induced by the RSA of the NPR occur \cite{XliS-23}. Considering the stimulated Raman scattering effect, the laser energy would be transferred from a 1030-nm pump wave to a ~1078-nm Stokes wave (downshifted by approximately 13 THz). Therefore, the front pulse preferentially undergoes energy downshifting, and the front pulse is subjected to strong RSA and evolves into a noise-like pulse (NLP) as shown in \textbf{Figure 5}.

The autocorrelation traces of five roundtrips randomly extracted from the pulse evolution process of the soliton compounds at cavity positions B, C, and after the filter are depicted at the bottom of Figure 5. The dynamic evolution of the autocorrelation traces as a unique property of the soliton compounds is demonstrated, which is consistent with the experimental observation.

\begin{figure}[htbp]
\centering\includegraphics[width=8.2 cm]{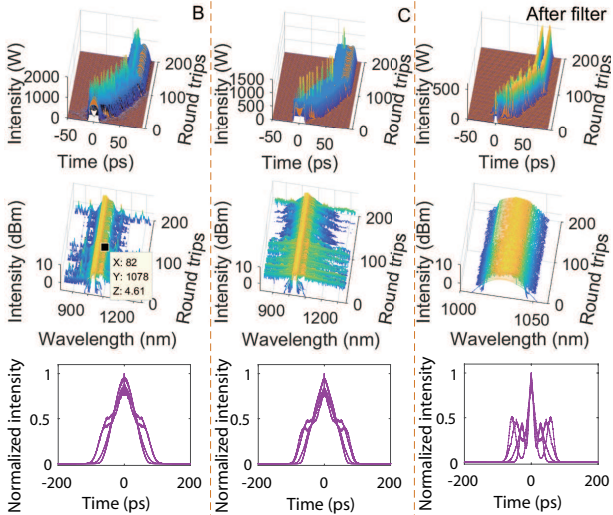} \caption{Soliton compound evolution in the time domain (up) and spectral domain (middle) at cavity positions B, C, and after the filter. The autocorrelation traces (down) of five roundtrips were randomly extracted.}\label{FL-Fig5}
\end{figure}

\subsection{NLP}
The length of $\rm{SMF_{1}}$ is further increased from 880 cm to 1480 cm, whereas that of $\rm{SMF_{2}}$ is decreased from 850 cm to 250 cm, in the PPE link with a fixed g$_0$ and SMF sum length (Simulation 4 in Table 1). As the pulse amplitude is further amplified when the pulse passes through the YDF, the peak power of the output pulse from the PPE link also increases, and can exceed 1200 W at position B, as shown in \textbf{Figure 6}. This value is greater than the CSP of the NPR, indicating that it is located in the RSA region, i.e. when the entire pulse passes through the NPR SA, the low-power portion undergoes less absorption, while the high-power one undergoes considerable absorption. The pulse constantly experiences the RSA effect while propagating in the cavity until the system reaches an equilibrium state; consequently, NLPs are formed without drastic nonlinear induced optical wave breaking.
\begin{figure}[htbp]
\centering\includegraphics[width=8.2 cm]{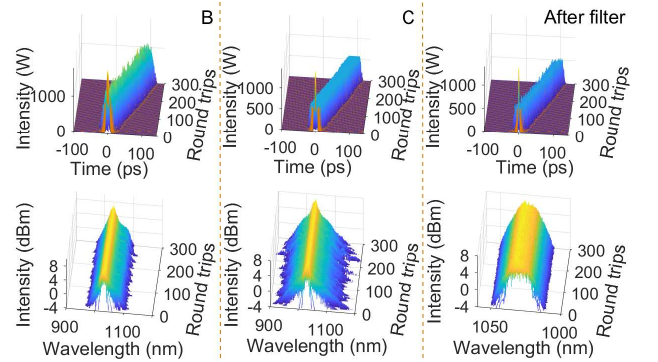} \caption{Calculated NLP evolution in time (up) and spectrum (down) domains at cavity position B, C and after filter.}\label{FL-Fig6}
\end{figure}

\section{Experimental results}\label{Sec4}
We construct fiber ring lasers, as shown in \textbf{Figure 7}, for the experiment. Briefly, they comprise an all-optical PPE link and a NPR mode locker. The PPE link includes an 8-nm BPF, a 976/1030-nm WDM, 850-mW LD, variable length SMF, 0.6-m YDF, 10\% OC, and another variable length SMF. The NPR includes two PCs used for finely adjusting the cavity intensity-dependent losses. While constructing the laser cavity, special care is taken to ensure that the total length of the cavity is 1790 cm. The SMF length between \rm{PC$_2$} and the BPF should be minimum; by adjusting the length of the SMF between points BPF and A (Figure 7) in experiments, the length of \rm{\emph{L}$_{CA}$} is made to correspond to the $\rm{SMF_{1}}$ length in the numerical simulation along the operating direction of the laser. By adjusting the length of the SMF between points B and C, length \rm{\emph{L}$_{BC}$} is made to correspond to the $\rm{SMF_{2}}$ length in simulation.
\begin{figure}[htbp]
\centering\includegraphics[width=7.5 cm]{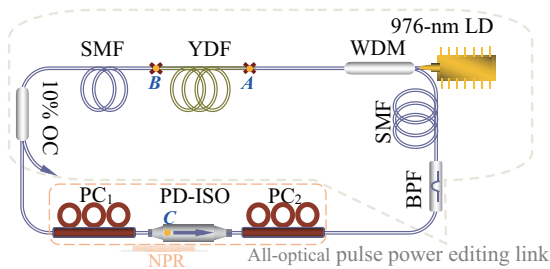} \caption{Schematic of the fiber soliton laser setup. LD: laser diode, WDM: wavelength division multiplexer, YDF: Yb-doped fiber, 10\% OC: 10\% output coupler, PC: polarization controller, PD-ISO: polarization-dependent isolator, NPR: nonlinear polarization rotation.}\label{FL-Fig7}
\end{figure}

Corresponding to the numerical simulation parameters, we perform experiments using four fiber lasers with uniform cavity lengths as listed in Table 1. The experimental results are summarized in \textbf{Table 2}.

\begin{table}[]
\caption{\label{tab:Table}%
Generated controllable pulse patterns in experiments. The fixed length sum of \rm{\emph{L}$_{CA}$}and \rm{\emph{L}$_{BC}$} is 1730 cm. }
\begin{tabular}{|l|l|l|l|l|}
\hline
\begin{tabular}[c]{@{}l@{}}$\rotatebox{270}{Results}$\end{tabular}& \begin{tabular}[c]{@{}l@{}}\rm{\emph{L}$_{BPF-A}$}\\(m)\end{tabular} & \begin{tabular}[c]{@{}l@{}}\rm{\emph{L}$_{BC}$}\\(m)\end{tabular} & \begin{tabular}[c]{@{}c@{}}Pump power\\(mW)\end{tabular} & \begin{tabular}[c]{@{}c@{}}Pulse patterns \\\textbf{Fiure 8}\end{tabular} \\ \hline
1   & 4.8                                             & 12.5                                             & 260.9-310.0
&\begin{tabular}[c]{@{}l@{}}Singe DS\\\textbf{Figure 8(a)}\end{tabular}  \\ \hline
2   & 6.8                                             & 10.5                                             & 268.3-302.8                                            &\begin{tabular}[c]{@{}l@{}}DS molecules\\\textbf{Figure 8(b)}\end{tabular} \\ \hline
3   & 8.8                                             & 8.5                                              & 270.2-322.3
&\begin{tabular}[c]{@{}l@{}}Soliton compounds\\\textbf{Figure 8(c)}\end{tabular}\\ \hline
4   & 14.8                                            & 2.5                                             & 282.5-650.4                                           &\begin{tabular}[c]{@{}l@{}}NLP\\\textbf{Figure 8(d)}\end{tabular}   \\ \hline
\end{tabular}
\end{table}

A single DS is generated when the pump power is 260.9 mW, as shown in \textbf{Figure 8}a. Further, we cut 2-m of SMF from \rm{\emph{L}$_{BC}$} of the DS cavity and fuse it with the \rm{\emph{L}$_{BPF-A}$} to rebuild another fiber laser. By carefully adjusting the PC states, DS molecules are obtained when the pump power is increased to 268.3 mW, as shown in Figure 8b.

It is important to note that in addition to DS molecules, other pulse patterns such as the DS and NLP can also be obtained by appropriately adjusting the pump power and the state of the PC in the same cavity. We can qualitatively understand these phenomena as follows: In addition to the discreet cavity structure design, the change in the nonlinear loss curve of the NPR mainly depends on the PC rotation angle. The dynamic interplay between the nonlinear loss of the NPR and other physical effects such as the gain saturation of the YDF, dispersion and Kerr nonlinearity of the fiber, and spectral filtering can also realize mode-locked operation, which is expected to generate pulse dynamic patterns at a constant repetition rate. Generally, single DS is easily produced if the NPR transmission curve has a high CSP \cite{XliS-23}. Assuming that the gain saturation of the gain fiber is sufficiently large, the pulse can be rapidly amplified in a very short gain fiber. Therefore, depending on the nonlinear effect of the fiber, pulse breaking occurs easily in an all-normal dispersion fiber laser \cite{GPAgrawal-26}. In this case, multiple pulse states, including DS molecules, are easily generated \cite{FLiP-41,WHRenningerA-42}. If the pump power is increased with a lower gain saturation, the pulse does not break. When the peak power of the pulse exceeds the CSP of the NPR transmission curve considerably, the pulse can easily evolve into an NLP depending on the RSA effect \cite{XliS-23}. Therefore, using pulse power editing along with NPR saturated absorption, and guided by numerical simulation, it is easier to achieve the controllable generation of pulse patterns such as the DS molecules displayed in Figure 8b.

2-m SMF is cut from \rm{\emph{L}$_{BC}$} of the DS molecule cavity and fused with \rm{\emph{L}$_{BPF-A}$} to rebuild the third fiber laser. The experimental results shown in Figure 8c confirm the existence of soliton compounds in the fiber laser, as depicted in Figure 5. In particular, the autocorrelation trace of the soliton compounds exhibits a dynamically variable shape as a function of the round-trip cavity period, as shown in Visualization 1; however, due to the superimposed recording of the autocorrelator, the autocorrelation trace in the experiment is recorded as the inset of Figure 8c, which is in good agreement with the numerical simulation (also see the numerical autocorrelation image at the bottom of Figure 5).

6-m SMF is further cut from \rm{\emph{L}$_{BC}$} of the soliton compound cavity and fused with \rm{\emph{L}$_{BPF-A}$} to rebuild the last fiber laser. As previously mentioned, the pulse power is edited largely into the NPR RSA region. On increasing the pump power to 282.5 mW, an NLP is produced. The signature autocorrelated signal, shown in the inset of Figure 8d with a narrow spike ¡°riding¡± on a wide pedestal is a clear indication of an NLP.

\begin{figure}[htbp]
\centering\includegraphics[width=8.3cm]{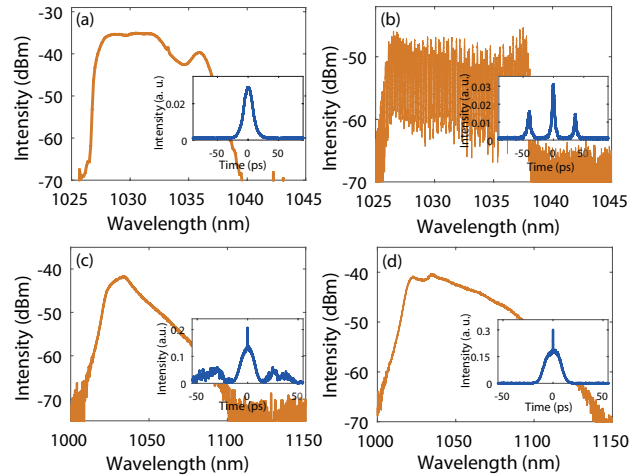} \caption{a) Optical spectrum and autocorrelation trace (inset) of DS. b) Optical spectrum and autocorrelation trace (inset) of the DS molecules. c) Optical spectrum and autocorrelation trace (inset) of the soliton compounds made up of DS and NLP  (For dynamic autocorrelation see Visualization 1 also). d) Optical spectrum and autocorrelation trace (inset) of the NLP.}\label{FL-Fig8}
\end{figure}

\section{Conclusion}\label{Sec5}
In conclusion, we presented a concise mechanism for the controllable generation of pulse patterns as predicted by theoretical works. The pulse power editing technique as well as the pivotal role of the NPR reverse saturable absorption provide new perspectives on complex pulse dynamics and inspire innovative cavity designs. Furthermore, by combining the pulse power editing technique with a saturated low-modulation depth absorber or a saturated absorber with a second mode-locking zone, the generation of pulse patterns is significant in nonlinear fiber optics. Moreover, the spatiotemporal dynamics of multimode optical solitons, and fields including hydrodynamics can benefit from the cross-fertilization of ideas.

\section*{Acknowledgements}
We acknowledge financial support in part by the National Natural Science Foundation of China (NSFC) under Grants 61605040 and 11374089, in part by the Natural Science Foundation of Hebei Province (NSFHP) under Grants F2017205162, F2017205060, and F2016205124, in part by the Program for High-Level Talents of Colleges and Universities in Hebei Province (PHLTCUHP) under Grant BJ2017020.


\end{document}